\documentclass[a4paper]{article}

\usepackage{INTERSPEECH2021}
\usepackage{mathtools}
\usepackage[table]{xcolor}
\usepackage{cite}
\definecolor{Gray}{gray}{0.85}

\title{The Sillwood Technologies System for the VoiceMOS Challenge 2022}
\name{Jiameng Gao$^1$}
\address{
  $^1$Sillwood Technologies, United Kingdom
}
\email{jiameng@sillwood.tech}

\begin{document}

\maketitle
\begin{abstract}
  In this paper we describe our entry for the VoiceMOS Challenge 2022 for both the main and out-of-domain (OOD) track of the competition. Our system is based on finetuning pre-trained self-supervised waveform prediction models, while improving its generalisation ability through stochastic weight averaging. Further, we use influence functions to identity possible low-quality data within the training set to further increase our model's performance for the OOD track. Our system ranked 5th and joint 7th for the main track and OOD track, respectively.
\end{abstract}

\section{Introduction}

Evaluation for speech synthesis has typically been a practically challenging task as many aspects of speech and its recorded waveforms contribute to its perceived naturalness. While many facets of speech, such as prosody \cite{klimkov2019finegrained, hodari2019intonation, morrison2020controllable, lancucki2021fastpitch}, energy \cite{raitio2020controllable, mohan2021ctrlp, ren2021fastspeech} and other features and representations \cite{henter2018deep, skerry2018towards, wang2018style, lee2019robust}, could be captured and recreated to a high fidelity \cite{shen2018natural, parallelwavegan, chen2021wavegrad, kong2021diffwave}, the sum of these parts do not yet account for the overall human perception of quality and fidelity in speech - a discrepancy to be explained by some undiscovered qualia. As such, the evaluation for speech synthesis and text-to-speech (TTS) systems rely on subjective human evaluation \cite{MOS}, meaning that unlike other areas in statistical and machine learning, the general "quality" and performance of speech synthesis systems could not be straightforwardly measured or optimised against using automated metrics. For TTS research, this means proper evaluation of speech synthesis systems could easily stretch into days or weeks, as opposed to hours or minutes, slowing down research cycles.

In this context, efforts have been made to automate the speech evaluation process, such as the ITU P. 563 algorithm \cite{itu563} and PESQ \cite{941023, itu862}, which were designed for measuring speech transmitted over communication channels such as the telephone; other metrics such cepstral distance, the loglikelihood ratio to an LPC model and the weighted spectral slope were researched as objective evaluation metrics for SPSS \cite{zen2009statistical} systems \cite{moller2010comparison, Remes2013ObjectiveEM, Norrenbrock2015QualityPO, ijima2016objective}.

An area related to quality estimation of synthetic speech is the use of generative adversarial network (GAN) setups for the training of speech synthesis systems, where generator and discriminator networks are trained jointly to improve the quality of acoustic models in TTS systems and/or neural vocoders \cite{melgan, gantts, parallelwavegan}. Here, the intended role of a discriminator network is to classify whether a waveform sequence is drawn from the data distribution or synthesised from the generator network.

The VoiceMOS challenge \cite{voicemospaper} is set up with the aim to develop highly accurate and correlative data-driven systems that can directly predict subjective evaluation scores, e.g. mean opinion score (MOS) \cite{MOS}, from synthesised speech, building on a growing library of work in this area \cite{falk2006single, 45744, yoshimura2016hierarchical, avila2019non, lo2021mosnet, cooper2021generalization, huang2021ldnet, zezario2021deep}. These systems are designed to be "single-ended" or "non-intrusive" such that generated speech can be evaluated without the need for gold-standard recordings. In this paper, we describe our entries for both tracks of the VoiceMOS challenge\footnote{Code: \texttt{https://github.com/sillwood/voicemos}}:
\begin{itemize}
  \item We follow \cite{cooper2021generalization} and build our systems for both tracks based on finetuning pre-trained self-supervised learning models for speech - namely wav2vec 2.0 \cite{baevski2020wav2vec}.
  \item We attempt to improve the generalisation capabilities of our models with stochastic weight averaging Gaussian \cite{izmailov2018averaging, maddox2019simple}.
  \item Furthermore, we filter the out-of-domain track training data with influence functions \cite{influencefunctions} with the aim to improve data quality.
\end{itemize}

\section{VoiceMOS challenge}


The VoiceMOS challenge \cite{voicemospaper} is made up of a main track and an out-of-domain (OOD) track, with the datasets for each track divided into training, development and testing sets. The training and development sets were released with information on the MOS score from each listener for each sample, some listener details and the MOS scores averaged across all listeners who had rated the sample. For both tracks, the test sets comprises of samples from systems, listeners and speakers unseen from the training data and therefore require some degree of generalisation in the MOS prediction systems.

The aim of the main track is to build highly accurate MOS prediction systems through utilising the 185 speech synthesis systems from previous Blizzard challenges, voice conversion challenges and other TTS samples \cite{hayashi2020espnet}; made up of both TTS and voice conversion systems \cite{cooper2021voices}. The main track dataset consists of multi-speaker data recorded in English, with the training set containing 4,973 audio samples; the development and test sets correspondingly contain 1,066 samples each.

The OOD track of the VoiceMOS challenge consists of a single speaker dataset with where, upon inspection, the synthesised language is Mandarin Chinese (with a Beijing accent). More challenging than the main track, the OOD track training set consists of 676 samples, of which only 136 have been labelled with corresponding listener and averaged MOS scores. Meanwhile, the development set for the OOD track also consists of 136 samples and the test set contains 540 samples. For our system, only the 136 labelled samples were used. Both tracks of the challenge consist of audio files (down)sampled at 16kHz.

\subsection{Evaluation}

Entries to the VoiceMOS challenge are evaluated with metrics across both the utterance and system level: mean squared error (MSE), Linear Correlation Coefficient (LCC), Spearman Rank Correlation Coefficient (SRCC), and Kendall Tau Rank Correlation (KTAU), with the system-level SRCC used as the metric to determine the main ranking of submitted systems.

\section{Our system}

\subsection{Finetuning self-supervised learning models for MOS prediction}


Through finetuning, pre-trained self-supervised learning (SSL) models \cite{devlin2019bert, NEURIPS2019_c04c19c2, brown2020language} have become a commonly used basis upon which high performance ML systems are built and its application to audio and speech on the waveform level have gained prominence within speech processing research \cite{oord2019representation, schneider2019wav2vec, hsu2021hubert, baevski2020wav2vec}.

\cite{cooper2021generalization} introduces SSL-MOS, an MOS prediction system built through finetuning SSL models such as HuBERT \cite{hsu2021hubert} and wav2vec 2.0 \cite{baevski2020wav2vec}, where they are used to generate time-averaged acoustic representations of speech samples, which are in turn projected onto MOS prediction scores using a single layer neural net, achieving high prediction accuracy and SRCC scores. It was found that finetuning the wav2vec 2.0 models, pre-trained on its base dataset, obtained the best results, with the wav2vec 2.0 Large model in particular performing best and significantly out-performing MOS prediction systems trained only on the task-specific BVCC dataset \cite{cooper2021voices}.

\subsection{SWA-Gaussian}


In order to improve on the generalisation ability of the SSL-MOS system, we leverage stochastic weight averaging Gaussian (SWAG). Stochastic weight averaging (SWA) \cite{izmailov2018averaging} is a model averaging method that is shown to improve model generalisation by building a model average as the model parameters $\theta$ moves around through stochastic gradient descent (SGD), assuming first that a local optimum has been found. For each $N$th sample $\theta_N$, the following average update rule is applied to obtain the SWA solution:
\begin{align}
    \theta_{\text{avg}, N} = \frac{\theta_{\text{avg}, N - 1} (N - 1) + \theta_{N}}{N}
    \label{eq:swa}
\end{align}
Where $\theta_{\text{avg}, N}$ is the averaged model parameters after $N$ samples. \cite{izmailov2018averaging} shows while SGD finds wider and flatter areas in the loss landscape, it does not always produce model parameters at, or close to, the centre of this space, whereas the SWA solution is able to lie much closer to the central region, and is therefore able to better generalise to test set data with minor additional computational costs.

\cite{maddox2019simple} extends SWA by mixing in Bayesian inference, wherein we attempt to obtain a posterior distribution $P(\theta | D)$ of the model parameters $\theta$ modelled as:
\begin{align}
    P(\theta | D) \propto P(D | \theta) P(\theta)
    \label{eq:bayes}
\end{align}
With $D$ being the (training) data, $P(D | \theta)$ the likelihood of the data under the model parameters $\theta$ and $P(\theta)$ a prior distribution over the weights. In the Bayesian setting, the probability of a new data point can be estimated by the marginal likelihood through Bayesian model averaging:
\begin{align}
    P(y_{\text{new}} | x_{\text{new}}, D) = \int P(y_{\text{new}} | x_{\text{new}}, \theta, D) P(\theta | D) d\theta
    \label{eq:bma}
\end{align}
However as modern neural network models require very high parameter counts, with even the small wav2vec 2.0 model containing 93 million trainable parameters, proper modelling of the posterior distribution and the marginal likelihood can be rather intractable. \cite{maddox2019simple} attempts to approximate the posterior distribution by introducing SWA-Gaussian (SWAG), which collects additional low-rank or diagonal covariance terms in a similar fashion to the SWA average in equation \ref{eq:swa}, enabling the construction of a Gaussian approximation distribution $Q(\theta | D)$ to $P(\theta | D)$ around the discovered local minima in the loss landscape.

Samples of parameters $\theta$ can then be obtained through Monte Carlo sampling on $Q(\theta | D)$ and in turn effect Bayesian model averaging:
\begin{align}
    P(y_{\text{new}} | x_{\text{new}}, D) &\approx \frac{1}{K}\sum_{k=1}^{K} P(y_{\text{new}} | \theta_k, x_{\text{new}}, D) \label{eq:swag_inf}\\
    \theta_k &\sim Q(\theta | D) = \mathcal{N}(\theta ; \mu_{\text{swag}}, \Sigma_{\text{swag}})
\end{align}
Where $\mu_{\text{swag}}$ and $\Sigma_{\text{swag}}$ are the SWAG mean and covariance terms collected. \cite{maddox2019simple} reports results showing SWAG to outperform other approaches in Bayesian deep learning.

\subsection{Influence functions for data debugging}

Meanwhile, \cite{influencefunctions} uses influence functions \cite{hampel1974influence} to show that despite the black-box nature of neural net models, we can still obtain information on the effect of training points on trained parameters of the network; more precisely, the influence $\mathcal{I}_{\text{up}, \text{params}} (z)$ of upweighting a training point $z = (x, y) \in D$ by a small amount $\epsilon$ on the trained model parameters $\hat{\theta}$ with respect to the loss $L$, where:
\begin{align}
    \hat{\theta}_{\epsilon, z} &\coloneqq \arg \min_\theta \big( L(D, \theta) + \epsilon L(z, \theta) \big)\\
    \text{s.t.} \quad \mathcal{I}_{\text{up}, \text{params}} (z) &\coloneqq \frac{d\hat{\theta}_{\epsilon, z}}{d\epsilon} \bigg|_{\epsilon=0}\\
    &= -H^{-1}_{\hat\theta} \nabla_\theta L(z, \hat\theta)\\
    \text{where} \quad H_{\hat\theta} &\coloneqq \nabla^{2}_{\theta} L (D, \hat\theta)
\end{align}
Here $H_{\hat\theta}$ is the Hessian evaluated at $\hat{\theta}$ and is assumed to be positive-definite.

Similarly, for a given test point $z_\text{test}$, we can calculate the influence of upweighting a training point $z$ on the loss at the test point $L(z_\text{test}, \hat\theta)$:
\begin{align}
    \mathcal{I}_{\text{up}, \text{loss}} (z, z_\text{test}) &\coloneqq \frac{dL(z_\text{test}, \hat{\theta}_{\epsilon, z})}{d\epsilon} \bigg|_{\epsilon=0}\\
    &= \nabla_\theta L(z_\text{test}, \hat{\theta})^\top \frac{d\hat{\theta}_{\epsilon, z}}{d\epsilon} \bigg|_{\epsilon=0} \\
    &= -\nabla_\theta L(z_\text{test}, \hat{\theta})^\top H^{-1}_{\hat\theta} \nabla_\theta L(z, \hat\theta)
\end{align}
Where a positive $\mathcal{I}_{\text{up}, \text{loss}} (z, z_\text{test})$ means a positive gradient in $L$. Here, \cite{influencefunctions} recognise that the Hessian-Vector Product $s_\text{test} \coloneqq H^{-1}_{\hat\theta} \nabla_\theta L(z_\text{test}, \hat\theta)$ can be implicitly approximated, pre-computed and stored, such that $\mathcal{I}_{\text{up}, \text{loss}} (z, z_\text{test}) = - s_\text{test}^\top \nabla_\theta L(z, \hat\theta)$ can be easily calculated for any training point $z$.

\cite{influencefunctions} shows that by obtaining influence values, we can identify and generate adversarial training examples to construct data-poisoning attacks. Furthermore, we can use influence values to debug datasets and identify "faulty" training data points, where a positive $-\mathcal{I}_{\text{up}, \text{loss}} (z, z_\text{test})$ indicates that the training point $z$ helps to reduce the loss on $z_\text{test}$, while negative values of $-\mathcal{I}_{\text{up}, \text{loss}} (z, z_\text{test})$ indicates a data point that exerts high negative influence on the test point $z_\text{test}$.

\subsection{Experiment} \label{sec:experiment}

For both tracks of the challenge, we follow the wav2vec 2.0 finetuning process set out in \cite{cooper2021generalization}, in which features from a wav2vec 2.0 model are extracted after the Transformer encoder layer, mean-pooled across time and projected down to a one-dimensional output for the score, trained with an L1 loss against the averaged MOS. A modification we make is in applying a sigmoid function to the output layer and scaling it to be $ \in[1, 5]$, though this may not have significant impact on model performance. We use the wav2vec 2.0 model from the \verb|fairseq| package \cite{ott2019fairseq} with the \verb|wav2vec_small| checkpoint, which is much smaller than the larger model (93 million parameters compared to 317 million) and showed similar performance after finetuning on the main track training set.

We train our models using PyTorch with automatic mixed precision training using the Brain Float 16 format \cite{bfloat16}, powered by an Nvidia RTX 3090 GPU. We first finetune our model for 30,000 iterations on the main track data for both tracks, using SGD with a learning rate of 0.001, momentum term of 0.9, cosine learning rate schedule ($T_{\text{max}} = 100$) as well as a batch-size of 8. The SWAG process is then run, where the system for each track is finetuned on its own training sets with a constant learning rate of 0.001 for SGD, a momentum value of 0.9. Due to the slight additional GPU memory requirement for SWAG, we use a batch-size of 4 with gradient accumulation every two steps, giving an effective batch-size of 8. We note that this is an incorrect application of SWAG for the OOD track, as the local minima has not settled into a local minima for the OOD track training data when the SWAG process is started. For checkpoint selection, we wait for the SWAG process to gather at least 5 samples before choosing the checkpoint that performs best on the system-level SRCC performance of the mean parameters of the SWAG distribution.

For OOD data debugging with influence functions, we generate features using the finetuned wav2vec 2.0 model and, similar to the Inception experiment in \cite{influencefunctions}, calculate the influence values for the final output layer only. We port the final layer parameters to and perform the influence function calculations in JAX in order to speed up the Hessian-Vector Product estimation. We restrict ourselves to the training set for $z_{\text{test}}$ (though this could also be applied to the development set) and identify $z_\text{test}$ that the model performs the worst on. We then calculate the $-\mathcal{I}_{\text{up}, \text{loss}}(z_i, z_\text{test})$ values for each training point $z_i \in D$ and keep a record of the most unhelpful data points (i.e. the most positive $-\mathcal{I}_{\text{up}, \text{loss}}(z, z_\text{test})$ values), from which we filter out two entries that were manually selected as having a higher averaged MOS than the audio quality suggests, namely \verb|sys2dea8-uttaca68d6.wav| and \verb|sys86a3b-utt81bb889.wav|; both files show significantly higher influence values ($10\times$ higher) compared to other training points in the dataset.

At inference, $K=10$ sets of parameters $\theta_k$ are drawn from the SWAG approximate distribution and the final model outputs are calculated as according to equation \ref{eq:swag_inf}.

Each configuration is trained across 5 different random seeds. For the systems we submitted to the VoiceMOS challenge, we select the models with the highest system-level SRCC scores on the development sets for each track; our entries ranked 5th and joint 7th for the main track and OOD track, respectively.

\subsection{Results} \label{results}

\begin{table}[]
    \caption{Test set results for main track systems. Bold fonts represent best mean scores but not necessarily statistically significant.}
    \footnotesize
    \centering
    \setlength{\tabcolsep}{2pt}
    \begin{tabular}{|c|cccc|}
        \hline
        \rowcolor{Gray}\multicolumn{5}{|c|}{System-level metrics} \\
        \hline
        Model & MSE & LCC & KTAU & SRCC \\
        \hline
        SSL-MOS & 0.105$\pm$0.007 & 0.937$\pm$0.005 & 0.784$\pm$0.010 & 0.934$\pm$0.005 \\
        SWAG & \textbf{0.102$\pm$0.017} & \textbf{0.939$\pm$0.004} & \textbf{0.788$\pm$0.009} & \textbf{0.936$\pm$0.004} \\
        \hline
        \rowcolor{Gray}\multicolumn{5}{|c|}{Utterance-level metrics} \\
        \hline
        Model & MSE & LCC & KTAU & SRCC \\
        \hline
        SSL-MOS & 0.252$\pm$0.012 & 0.871$\pm$0.002 & 0.692$\pm$0.003 & 0.867$\pm$0.002 \\
        SWAG & \textbf{0.247$\pm$0.026} & \textbf{0.876$\pm$0.003} & \textbf{0.698$\pm$0.003} & \textbf{0.872$\pm$0.003} \\
        \hline
    \end{tabular}
    \label{tab:main_track}
\end{table}

The results for the main track test set data are shown in table \ref{tab:main_track}, where the SWAG model performs marginally better than the vanilla SSL-MOS model for both the system-level and utterance-level metrics. However this pattern is more mixed for the OOD track, as shown in table \ref{tab:ood_track}, where the SWAG model is generally worse compared to the SSL-MOS model, while the SWAG+Inf model performs better than both on the utterance-level metrics, and worse than the SSL-MOS model for system-level KTAU and SRCC scores.

\begin{table}[]
    \caption{Test set results for OOD track data. SWAG+Inf denote models trained with SWAG on data filtered using influence values. Note that SWAG here may have been incorrectly applied (Section \ref{sec:experiment}). Bold fonts represent best mean scores but not necessarily statistically significant.}
    \footnotesize
    \centering
    \setlength{\tabcolsep}{2pt}
    \begin{tabular}{|c|cccc|}
        \hline
        \rowcolor{Gray}\multicolumn{5}{|c|}{System-level metrics} \\
        \hline
        System & MSE & LCC & KTAU & SRCC \\
        \hline
        SSL-MOS & 0.162$\pm$0.052 & 0.959$\pm$0.004 & \textbf{0.876$\pm$0.020} & \textbf{0.970$\pm$0.008} \\
        SWAG & 0.177$\pm$0.105 & 0.962$\pm$0.008 & 0.862$\pm$0.018 & 0.962$\pm$0.010 \\
        SWAG+Inf & \textbf{0.156$\pm$0.028} & \textbf{0.964$\pm$0.004} & 0.855$\pm$0.012 & 0.957$\pm$0.006 \\
        \hline
        \rowcolor{Gray}\multicolumn{5}{|c|}{Utterance-level metrics} \\
        \hline
        System & MSE & LCC & KTAU & SRCC \\
        \hline
        SSL-MOS & 0.308$\pm$0.053 & 0.884$\pm$0.007 & 0.672$\pm$0.013 & 0.855$\pm$0.011 \\
        SWAG & 0.317$\pm$0.102 & 0.887$\pm$0.006 & \textbf{0.674$\pm$0.008} & 0.856$\pm$0.008 \\
        SWAG+Inf & \textbf{0.298$\pm$0.031} & \textbf{0.890$\pm$0.008} & \textbf{0.674$\pm$0.013} & \textbf{0.857$\pm$0.011} \\
        \hline
    \end{tabular}
    \label{tab:ood_track}
\end{table}

\begin{figure}[t]
  \centering
   \includegraphics[clip, trim=1.5cm 0cm 2cm 1cm, width=1.00\linewidth]{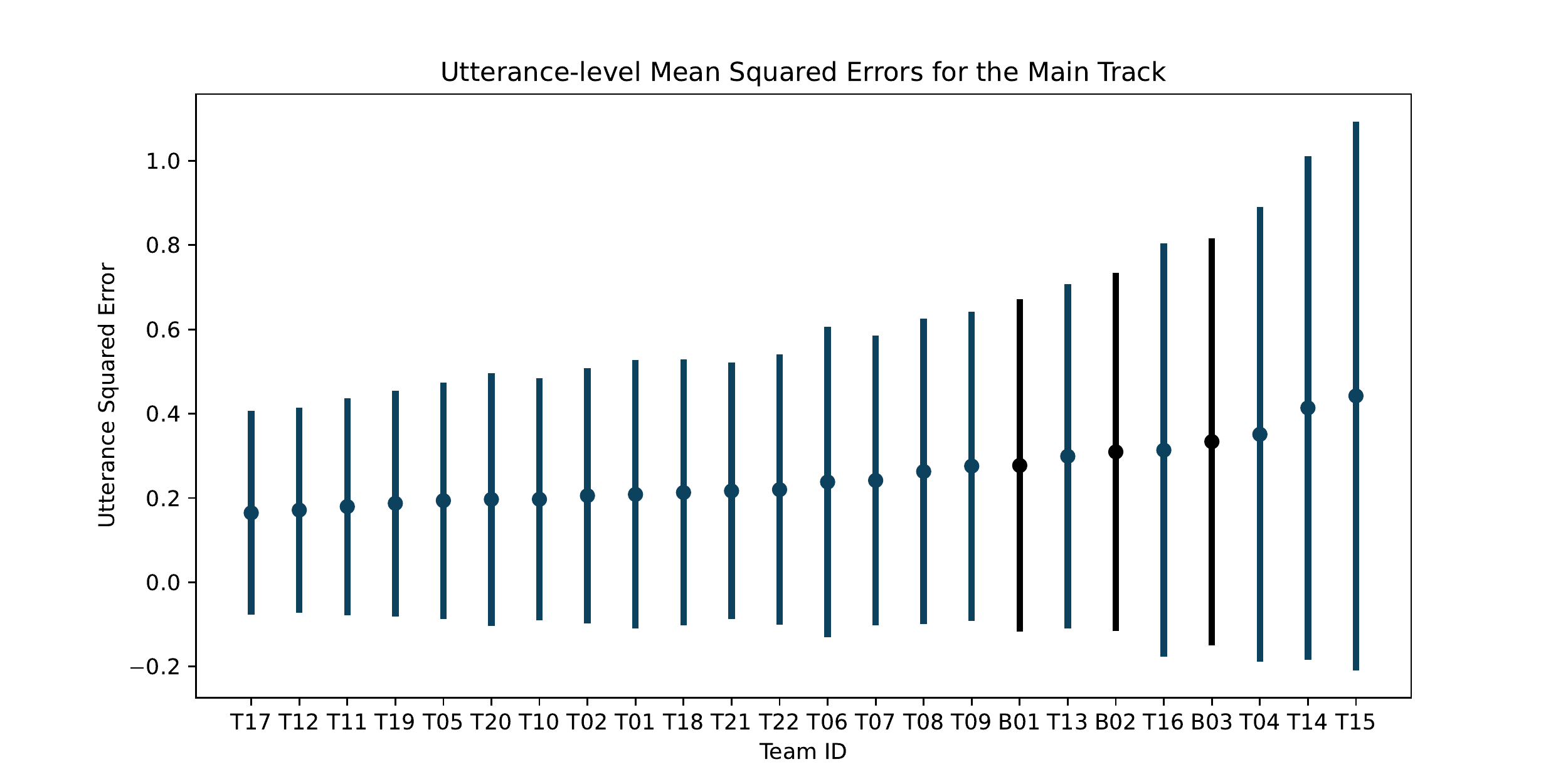}
  \caption{MSE scores across all teams for the main track of the VoiceMOS challenge, our team code is T07.}
  \label{fig:main_mse}
\end{figure}

\begin{figure}[t]
  \centering
   \includegraphics[clip, trim=1.5cm 0cm 2cm 1cm, width=1.00\linewidth]{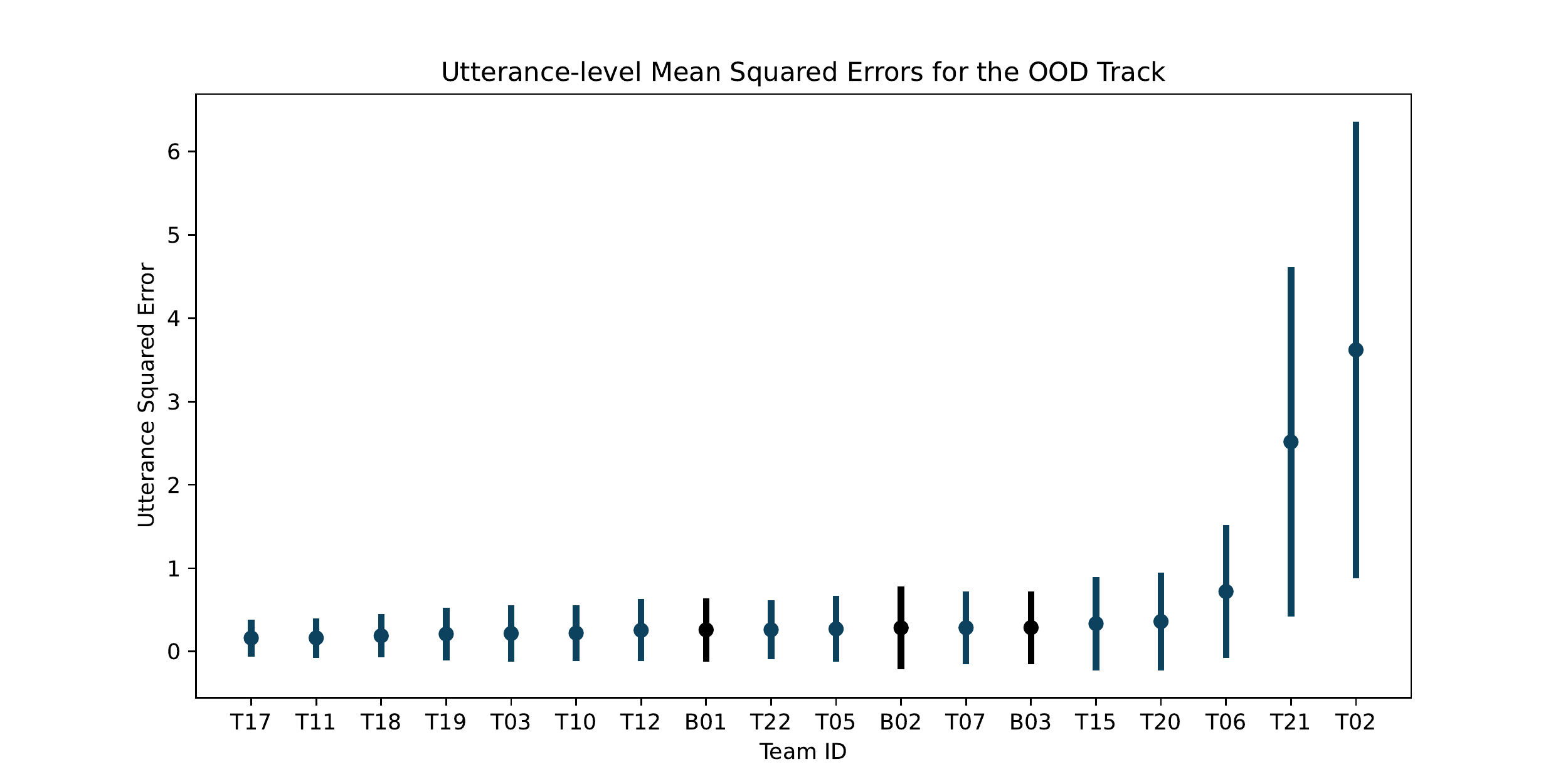}
  \caption{MSE scores across all teams for the OOD track of the VoiceMOS challenge, our team code is T07.}
  \label{fig:ood_mse}
\end{figure}

Our models rank much lower in the utterance-level MSE than other systems in the VoiceMOS challenge, as shown in figures \ref{fig:main_mse} and \ref{fig:ood_mse}, compared to the system SRCC score ranking; upon investigation, we found that our models for both tracks tend to over-predict MOS compared to the ground-truth.

\section{Discussion}


As shown in Section \ref{results}, while the addition of SWAG can positively impact model performance for the main track dataset, its effect on the OOD track data is somewhat mixed, though this may be due to the fact that the OOD track SWAG was incorrectly applied. Meanwhile, the use of influence functions to filter out data generally improves on the SWAG model and can outperform the SSL-MOS model on many of the metrics in the OOD track.

Nonetheless, the variance in the metrics, effected through different random seeds, makes it difficult to assert that the modifications introduced in this paper truly improves on the performance of the original SSL-MOS model, as the differences are comparatively minor. Furthermore, such variances could also mean the difference between a top-ranking system in the VoiceMOS challenge and one that is ranked much lower.


An aspect that our system does not address is the ability to accommodate multiple and different sample-rates in input audio samples, which has a notable impact on audio quality. Many modern TTS systems use vocoders that generate speech at sample rates higher than the 16kHz samples used for the VoiceMOS challenge \cite{shen2018natural, kalchbrenner2018efficient, parallelwavegan, chen2021wavegrad}, meaning that many systems used in real-world applications are not catered for by the base SSL-MOS system, which is trained with 16kHz data. It would be interesting to build waveform-level speech processing systems that are agnostic or invariant to sampling rates. One route could be to create systems that can operating on some continuous-time representation of a waveform instead of discrete-time samples \cite{Karras2021, Gordon:2020:Convolutional_Conditional_Neural_Processes}.


Lastly, the absolute value of MOS scores may not be a sustainable metric to use in terms of reproduciblity. As MOS is an absolute score assigned by subjective listeners, the MOS for a specific speech sample or system may change across listening experiments or decrease (or "drift") over time as the public expectations for synthesised speech increase. Evaluation metrics such as MUSHRA \cite{itu1534} tries to encapsulate this through rating synthesis systems relative to a reference and baseline sample, in order to ensure some form of continuity across experiments. Along these lines, it may be interesting to leverage results from other subjective evaluation methods, such as MUSHRA experiments or A/B tests, to create systems that predict rankings of speech synthesis systems instead raw MOS values.

\section{Acknowledgements}

We would like to thank the VoiceMOS challenge team for the organisation of the challenge and their helpful correspondence throughout the competition.

This paper as well as code create for our experiments were typed on a Filco Ninja Majestouch-2 keyboard.

\bibliographystyle{IEEEtran}

\bibliography{mybib}

\begin{thebibliography}{10}
\providecommand{\url}[1]{#1}
\csname url@samestyle\endcsname
\providecommand{\newblock}{\relax}
\providecommand{\bibinfo}[2]{#2}
\providecommand{\BIBentrySTDinterwordspacing}{\spaceskip=0pt\relax}
\providecommand{\BIBentryALTinterwordstretchfactor}{4}
\providecommand{\BIBentryALTinterwordspacing}{\spaceskip=\fontdimen2\font plus
\BIBentryALTinterwordstretchfactor\fontdimen3\font minus
  \fontdimen4\font\relax}
\providecommand{\BIBforeignlanguage}[2]{{%
\expandafter\ifx\csname l@#1\endcsname\relax
\typeout{** WARNING: IEEEtran.bst: No hyphenation pattern has been}%
\typeout{** loaded for the language `#1'. Using the pattern for}%
\typeout{** the default language instead.}%
\else
\language=\csname l@#1\endcsname
\fi
#2}}
\providecommand{\BIBdecl}{\relax}
\BIBdecl

\bibitem{klimkov2019finegrained}
V.~Klimkov, S.~Ronanki, J.~Rohnke, and T.~Drugman, ``Fine-grained robust
  prosody transfer for single-speaker neural text-to-speech,'' 2019.

\bibitem{hodari2019intonation}
Z.~Hodari, O.~Watts, and S.~King, ``Using generative modelling to produce
  varied intonation for speech synthesis,'' \emph{10th ISCA Workshop on Speech
  Synthesis (SSW 10)}, Sep 2019.

\bibitem{morrison2020controllable}
M.~Morrison, Z.~Jin, J.~Salamon, N.~J. Bryan, and G.~J. Mysore, ``Controllable
  neural prosody synthesis,'' in \emph{Interspeech 2020}, 2020.

\bibitem{lancucki2021fastpitch}
A.~{\L}a{\'n}cucki, ``Fastpitch: Parallel text-to-speech with pitch
  prediction,'' in \emph{ICASSP 2021-2021 IEEE International Conference on
  Acoustics, Speech and Signal Processing (ICASSP)}.\hskip 1em plus 0.5em minus
  0.4em\relax IEEE, 2021, pp. 6588--6592.

\bibitem{raitio2020controllable}
T.~Raitio, R.~Rasipuram, and D.~Castellani, ``Controllable neural
  text-to-speech synthesis using intuitive prosodic features,'' in
  \emph{Interspeech 2020}, 2020.

\bibitem{mohan2021ctrlp}
D.~S.~R. Mohan, V.~Hu, T.~H. Teh, A.~Torresquintero, C.~G.~R. Wallis, M.~Staib,
  L.~Foglianti, J.~Gao, and S.~King, ``{Ctrl-P: Temporal Control of Prosodic
  Variation for Speech Synthesis},'' in \emph{Interspeech 2021}, 2021.

\bibitem{ren2021fastspeech}
Y.~Ren, C.~Hu, X.~Tan, T.~Qin, S.~Zhao, Z.~Zhao, and T.-Y. Liu, ``Fastspeech 2:
  Fast and high-quality end-to-end text to speech,'' in \emph{International
  Conference on Learning Representations}, 2021.

\bibitem{henter2018deep}
G.~E. Henter, J.~Lorenzo-Trueba, X.~Wang, and J.~Yamagishi, ``Deep
  encoder-decoder models for unsupervised learning of controllable speech
  synthesis,'' \emph{arXiv preprint arXiv:1807.11470}, 2018.

\bibitem{skerry2018towards}
R.~Skerry-Ryan, E.~Battenberg, Y.~Xiao, Y.~Wang, D.~Stanton, J.~Shor, R.~Weiss,
  R.~Clark, and R.~A. Saurous, ``Towards end-to-end prosody transfer for
  expressive speech synthesis with tacotron,'' in \emph{international
  conference on machine learning}.\hskip 1em plus 0.5em minus 0.4em\relax PMLR,
  2018, pp. 4693--4702.

\bibitem{wang2018style}
Y.~Wang, D.~Stanton, Y.~Zhang, R.-S. Ryan, E.~Battenberg, J.~Shor, Y.~Xiao,
  Y.~Jia, F.~Ren, and R.~A. Saurous, ``Style tokens: Unsupervised style
  modeling, control and transfer in end-to-end speech synthesis,'' in
  \emph{International Conference on Machine Learning}.\hskip 1em plus 0.5em
  minus 0.4em\relax PMLR, 2018, pp. 5180--5189.

\bibitem{lee2019robust}
Y.~Lee and T.~Kim, ``Robust and fine-grained prosody control of end-to-end
  speech synthesis,'' in \emph{ICASSP 2019-2019 IEEE International Conference
  on Acoustics, Speech and Signal Processing (ICASSP)}.\hskip 1em plus 0.5em
  minus 0.4em\relax IEEE, 2019, pp. 5911--5915.

\bibitem{shen2018natural}
J.~Shen, R.~Pang, R.~J. Weiss, M.~Schuster, N.~Jaitly, Z.~Yang, Z.~Chen,
  Y.~Zhang, Y.~Wang, R.~Skerrv-Ryan \emph{et~al.}, ``Natural tts synthesis by
  conditioning wavenet on mel spectrogram predictions,'' in \emph{2018 IEEE
  international conference on acoustics, speech and signal processing
  (ICASSP)}.\hskip 1em plus 0.5em minus 0.4em\relax IEEE, 2018, pp. 4779--4783.

\bibitem{parallelwavegan}
R.~Yamamoto, E.~Song, and J.-M. Kim, ``Parallel wavegan: A fast waveform
  generation model based on generative adversarial networks with
  multi-resolution spectrogram,'' in \emph{ICASSP 2020}, 2020.

\bibitem{chen2021wavegrad}
N.~Chen, Y.~Zhang, H.~Zen, R.~J. Weiss, M.~Norouzi, and W.~Chan, ``Wavegrad:
  Estimating gradients for waveform generation,'' in \emph{International
  Conference on Learning Representations}, 2021.

\bibitem{kong2021diffwave}
Z.~Kong, W.~Ping, J.~Huang, K.~Zhao, and B.~Catanzaro, ``Diffwave: A versatile
  diffusion model for audio synthesis,'' in \emph{International Conference on
  Learning Representations}, 2021.

\bibitem{MOS}
{ITU-T Rec. P.85}, ``A method for subjective performance assessment of the
  quality of speech voice output devices,'' Int. Telecomm. Union, Geneva, 1994.

\bibitem{itu563}
{ITU-T Rec. P.563}, ``{Single-ended method for objective speech quality
  assessment in narrow-band telephony applications},'' Int. Telecomm. Union,
  Geneva, 2004.

\bibitem{941023}
A.~Rix, J.~Beerends, M.~Hollier, and A.~Hekstra, ``{Perceptual evaluation of
  speech quality (PESQ)-a new method for speech quality assessment of telephone
  networks and codecs},'' in \emph{2001 IEEE International Conference on
  Acoustics, Speech, and Signal Processing. Proceedings (Cat. No.01CH37221)},
  2001.

\bibitem{itu862}
{ITU-T Rec. P.862}, ``{Perceptual evaluation of speech quality (PESQ): An
  objective method for end-to-end speech quality assessment of narrow-band
  telephone networks and speech codecs},'' Int. Telecomm. Union, Geneva, 2001.

\bibitem{zen2009statistical}
H.~Zen, K.~Tokuda, and A.~W. Black, ``Statistical parametric speech
  synthesis,'' \emph{speech communication}, 2009.

\bibitem{moller2010comparison}
S.~M{\"o}ller, F.~Hinterleitner, T.~H. Falk, and T.~Polzehl, ``Comparison of
  approaches for instrumentally predicting the quality of text-to-speech
  systems,'' in \emph{Eleventh Annual Conference of the International Speech
  Communication Association}, 2010.

\bibitem{Remes2013ObjectiveEM}
U.~Remes, R.~Karhila, and M.~Kurimo, ``{Objective evaluation measures for
  speaker-adaptive HMM-TTS systems},'' in \emph{SSW}, 2013.

\bibitem{Norrenbrock2015QualityPO}
C.~Norrenbrock, F.~Hinterleitner, U.~Heute, and S.~M{\"o}ller, ``Quality
  prediction of synthesized speech based on perceptual quality dimensions,''
  \emph{Speech Commun.}, vol.~66, pp. 17--35, 2015.

\bibitem{ijima2016objective}
Y.~Ijima, T.~Asami, and H.~Mizuno, ``Objective evaluation using association
  between dimensions within spectral features for statistical parametric speech
  synthesis.'' in \emph{INTERSPEECH}, 2016, pp. 337--341.

\bibitem{melgan}
K.~Kumar, R.~Kumar, T.~de~Boissiere, L.~Gestin, W.~Z. Teoh, J.~Sotelo,
  A.~de~Br\'{e}bisson, Y.~Bengio, and A.~C. Courville, ``Melgan: Generative
  adversarial networks for conditional waveform synthesis,'' in \emph{Advances
  in Neural Information Processing Systems}, vol.~32.\hskip 1em plus 0.5em
  minus 0.4em\relax Curran Associates, Inc., 2019.

\bibitem{gantts}
M.~Bińkowski, J.~Donahue, S.~Dieleman, A.~Clark, E.~Elsen, N.~Casagrande,
  L.~C. Cobo, and K.~Simonyan, ``High fidelity speech synthesis with
  adversarial networks,'' in \emph{International Conference on Learning
  Representations}, 2020.

\bibitem{voicemospaper}
\BIBentryALTinterwordspacing
W.-C. Huang, E.~Cooper, Y.~Tsao, H.-M. Wang, T.~Toda, and J.~Yamagishi, ``{The
  VoiceMOS Challenge 2022},'' 2022. [Online]. Available:
  \url{https://arxiv.org/abs/2203.11389}
\BIBentrySTDinterwordspacing

\bibitem{falk2006single}
T.~H. Falk and W.-Y. Chan, ``Single-ended speech quality measurement using
  machine learning methods,'' \emph{IEEE Transactions on Audio, Speech, and
  Language Processing}, vol.~14, no.~6, 2006.

\bibitem{45744}
B.~Patton, Y.~Agiomyrgiannakis, M.~Terry, K.~Wilson, R.~A. Saurous, and
  D.~Sculley, ``{AutoMOS: Learning a non-intrusive assessor of
  naturalness-of-speech},'' in \emph{NIPS 2016 End-to-end Learning for Speech
  and Audio Processing Workshop}, 2016.

\bibitem{yoshimura2016hierarchical}
T.~Yoshimura, G.~E. Henter, O.~Watts, M.~Wester, J.~Yamagishi, and K.~Tokuda,
  ``A hierarchical predictor of synthetic speech naturalness using neural
  networks,'' in \emph{Interspeech}, 2016, pp. 342--346.

\bibitem{avila2019non}
A.~R. Avila, H.~Gamper, C.~Reddy, R.~Cutler, I.~Tashev, and J.~Gehrke,
  ``Non-intrusive speech quality assessment using neural networks,'' in
  \emph{ICASSP 2019-2019 IEEE International Conference on Acoustics, Speech and
  Signal Processing (ICASSP)}.\hskip 1em plus 0.5em minus 0.4em\relax IEEE,
  2019.

\bibitem{lo2021mosnet}
C.-C. Lo, S.-W. Fu, W.-C. Huang, X.~Wang, J.~Yamagishi, Y.~Tsao, and H.-M.
  Wang, ``{MOSNet: Deep Learning based Objective Assessment for Voice
  Conversion},'' in \emph{Interspeech 2020}, 2021.

\bibitem{cooper2021generalization}
E.~Cooper, W.-C. Huang, T.~Toda, and J.~Yamagishi, ``{Generalization ability of
  MOS prediction networks},'' \emph{arXiv preprint arXiv:2110.02635}, 2021.

\bibitem{huang2021ldnet}
W.-C. Huang, E.~Cooper, J.~Yamagishi, and T.~Toda, ``{LDNet: Unified Listener
  Dependent Modeling in MOS Prediction for Synthetic Speech},'' \emph{arXiv
  preprint arXiv:2110.09103}, 2021.

\bibitem{zezario2021deep}
R.~E. Zezario, S.-W. Fu, F.~Chen, C.-S. Fuh, H.-M. Wang, and Y.~Tsao, ``Deep
  learning-based non-intrusive multi-objective speech assessment model with
  cross-domain features,'' \emph{arXiv preprint arXiv:2111.02363}, 2021.

\bibitem{baevski2020wav2vec}
A.~Baevski, Y.~Zhou, A.~Mohamed, and M.~Auli, ``wav2vec 2.0: A framework for
  self-supervised learning of speech representations,'' \emph{Advances in
  Neural Information Processing Systems}, vol.~33, pp. 12\,449--12\,460, 2020.

\bibitem{izmailov2018averaging}
P.~Izmailov, D.~Podoprikhin, T.~Garipov, D.~Vetrov, and A.~G. Wilson,
  ``Averaging weights leads to wider optima and better generalization,''
  \emph{arXiv preprint arXiv:1803.05407}, 2018.

\bibitem{maddox2019simple}
W.~J. Maddox, P.~Izmailov, T.~Garipov, D.~P. Vetrov, and A.~G. Wilson, ``A
  simple baseline for bayesian uncertainty in deep learning,'' \emph{Advances
  in Neural Information Processing Systems}, vol.~32, 2019.

\bibitem{influencefunctions}
P.~W. Koh and P.~Liang, ``Understanding black-box predictions via influence
  functions,'' in \emph{Proceedings of the 34th International Conference on
  Machine Learning}.\hskip 1em plus 0.5em minus 0.4em\relax PMLR, 2017.

\bibitem{hayashi2020espnet}
T.~Hayashi, R.~Yamamoto, K.~Inoue, T.~Yoshimura, S.~Watanabe, T.~Toda,
  K.~Takeda, Y.~Zhang, and X.~Tan, ``Espnet-tts: Unified, reproducible, and
  integratable open source end-to-end text-to-speech toolkit,'' in \emph{ICASSP
  2020-2020 IEEE international conference on acoustics, speech and signal
  processing (ICASSP)}.\hskip 1em plus 0.5em minus 0.4em\relax IEEE, 2020, pp.
  7654--7658.

\bibitem{cooper2021voices}
E.~Cooper and J.~Yamagishi, ``How do voices from past speech synthesis
  challenges compare today?'' \emph{Speech Synthesis Workshop}, 2021.

\bibitem{devlin2019bert}
J.~Devlin, M.-W. Chang, K.~Lee, and K.~Toutanova, ``{BERT: Pre-training of Deep
  Bidirectional Transformers for Language Understanding},'' 2019.

\bibitem{NEURIPS2019_c04c19c2}
A.~Conneau and G.~Lample, ``Cross-lingual language model pretraining,'' in
  \emph{Advances in Neural Information Processing Systems}, H.~Wallach,
  H.~Larochelle, A.~Beygelzimer, F.~d\textquotesingle Alch\'{e}-Buc, E.~Fox,
  and R.~Garnett, Eds., vol.~32.\hskip 1em plus 0.5em minus 0.4em\relax Curran
  Associates, Inc., 2019.

\bibitem{brown2020language}
T.~Brown, B.~Mann, N.~Ryder, M.~Subbiah, J.~D. Kaplan, P.~Dhariwal,
  A.~Neelakantan, P.~Shyam, G.~Sastry, A.~Askell \emph{et~al.}, ``Language
  models are few-shot learners,'' \emph{Advances in neural information
  processing systems}, vol.~33, pp. 1877--1901, 2020.

\bibitem{oord2019representation}
A.~van~den Oord, Y.~Li, and O.~Vinyals, ``Representation learning with
  contrastive predictive coding,'' 2019.

\bibitem{schneider2019wav2vec}
S.~Schneider, A.~Baevski, R.~Collobert, and M.~Auli, ``wav2vec: Unsupervised
  pre-training for speech recognition,'' \emph{arXiv preprint
  arXiv:1904.05862}, 2019.

\bibitem{hsu2021hubert}
W.-N. Hsu, B.~Bolte, Y.-H.~H. Tsai, K.~Lakhotia, R.~Salakhutdinov, and
  A.~Mohamed, ``Hubert: Self-supervised speech representation learning by
  masked prediction of hidden units,'' \emph{IEEE/ACM Transactions on Audio,
  Speech, and Language Processing}, vol.~29, pp. 3451--3460, 2021.

\bibitem{hampel1974influence}
F.~R. Hampel, ``The influence curve and its role in robust estimation,''
  \emph{Journal of the american statistical association}, vol.~69, no. 346, pp.
  383--393, 1974.

\bibitem{ott2019fairseq}
M.~Ott, S.~Edunov, A.~Baevski, A.~Fan, S.~Gross, N.~Ng, D.~Grangier, and
  M.~Auli, ``fairseq: A fast, extensible toolkit for sequence modeling,'' in
  \emph{Proceedings of NAACL-HLT 2019: Demonstrations}, 2019.

\bibitem{bfloat16}
\BIBentryALTinterwordspacing
``{The BFLOAT16 Numerical Format},'' {Accessed: 2022}. [Online]. Available:
  \url{https://cloud.google.com/tpu/docs/bfloat16}
\BIBentrySTDinterwordspacing

\bibitem{kalchbrenner2018efficient}
N.~Kalchbrenner, E.~Elsen, K.~Simonyan, S.~Noury, N.~Casagrande, E.~Lockhart,
  F.~Stimberg, A.~Oord, S.~Dieleman, and K.~Kavukcuoglu, ``Efficient neural
  audio synthesis,'' in \emph{International Conference on Machine
  Learning}.\hskip 1em plus 0.5em minus 0.4em\relax PMLR, 2018, pp. 2410--2419.

\bibitem{Karras2021}
T.~Karras, M.~Aittala, S.~Laine, E.~H\"ark\"onen, J.~Hellsten, J.~Lehtinen, and
  T.~Aila, ``Alias-free generative adversarial networks,'' in \emph{Proc.
  NeurIPS}, 2021.

\bibitem{Gordon:2020:Convolutional_Conditional_Neural_Processes}
\BIBentryALTinterwordspacing
J.~Gordon, W.~P. Bruinsma, A.~Y.~K. Foong, J.~Requeima, Y.~Dubois, and R.~E.
  Turner, ``Convolutional conditional neural processes,'' in
  \emph{International Conference on Learning Representations}, 2020. [Online].
  Available: \url{https://openreview.net/forum?id=Skey4eBYPS}
\BIBentrySTDinterwordspacing

\bibitem{itu1534}
{ITU-R BS.1534}, ``{Method for the subjective assessment of intermediate
  quality level of audio systems},'' Int. Telecomm. Union, Geneva, 2003.

\end{thebibliography}


\end{document}